\documentclass[12pt,draftcls,onecolumn]{IEEEtran}
\IEEEoverridecommandlockouts                              	
\usepackage[T1]{fontenc}
\usepackage{graphicx}
\graphicspath{{figures/}}
\usepackage{amssymb,amsmath,amsfonts}
\usepackage{url,changebar,bm,xspace,dsfont}
\usepackage{theorem}
\usepackage{epsfig}
\usepackage{epstopdf}
\usepackage{cite}
\usepackage{verbatim}
\usepackage[usenames,dvipsnames]{color}
\usepackage{flushend}
\usepackage{tikz}

\newtheorem{remark}{Remark}
\newtheorem{definition}{\bfseries Definition}
\newtheorem{theorem}{\bfseries Theorem}

\newtheorem{assump}{\bfseries Assumption}

\newtheorem{proposition}{\bfseries Proposition}

\def\bnull{\bm{0}}

\def\w{\bm{w}}\def\x{\bm{x}}\def\y{\bm{y}}\def\a{\bm{a}}\def\v{\bm{v}}\def\f{\bm{f}}\def\z{\bm{z}}\def\g{\bm{g}}

\def\bphi{\bm{\varphi}}\def\bph{\bm{\phi}}

\begin{document}

\title{Exponential Stability of Homogeneous Positive Systems of Degree One With Time-Varying Delays}

\author{Hamid Reza Feyzmahdavian, Themistoklis Charalambous, and Mikael Johansson
\thanks{H. R. Feyzmahdavian, T. Charalambous, and M. Johansson are with ACCESS Linnaeus Center, School of Electrical
Engineering, KTH-Royal Institute of Technology, Stockholm, Sweden.
Emails: {\tt \{hamidrez, themisc, mikaelj\}@kth.se.}}
}

\maketitle

%
%

\begin{abstract}

While the asymptotic stability of positive linear systems in the presence of bounded time delays has been thoroughly investigated, the theory for nonlinear positive systems is considerably less well-developed. This paper presents a set of conditions for establishing delay-independent stability and bounding the decay rate of a significant class of nonlinear positive systems which includes positive linear systems as a special case. Specifically, when the time delays have a known upper bound, we derive necessary and sufficient conditions for exponential stability of (a) continuous-time positive systems whose vector fields are homogeneous and cooperative, and (b) discrete-time positive systems whose vector fields are homogeneous and order preserving. We then present explicit expressions that allow us to quantify the impact of delays on the decay rate and show that the best decay rate of positive linear systems that our bounds provide can be found via convex optimization. Finally, we extend the results to general linear systems with time-varying delays.

\end{abstract}

%
%

\section{Introduction}\label{sec:intro}

Positive systems are dynamical systems whose state variables are constrained to be nonnegative for all time whenever the initial conditions are nonnegative~\cite{Farina:00}. Due to their importance and wide applicability, the analysis  and control of positive systems has attracted considerable attention from the control community (see, \emph{e.g.},~\cite{Leenheer:01,2009:Florian,Rantzer:11,Briat:11,Tanaka:11,Feyzmahdavian:13,Feyzmahdavian:13-1} and references therein).

Since time delays are omnipresent in engineering systems, the study of stability and control of dynamical systems with delayed states is essential and of practical importance. For general systems, the existence of time delays may impair performance, induce oscillations and even instability~\cite{Driver:62}. In contrast, positive linear systems have been shown to be insensitive to certain classes of time delays in the sense that a positive linear system is asymptotically stable for all bounded delays if and only if the corresponding delay-free system is asymptotically stable~\cite{Haddad:04,Rami:09,Liu:10,Liu:09}. In addition, if a positive linear system is asymptotically stable for an arbitrary constant delay and some positive initial conditions, the delay-free system is globally asymptotically stable~\cite{Rami:09}.

Many important positive systems are \textit{nonlinear}. It is thus natural to ask if the insensitivity properties of positive linear systems with respect to time delays will hold also for nonlinear positive systems. In~\cite{Mason:09}, it was shown that for a particular class of nonlinear positive systems, \textit{homogeneous cooperative} systems with \textit{constant} delays, this is indeed the case. It is clear that constant delays is an idealized assumption as time delays are often \textit{time-varying} in practice. However, to the best of our knowledge, there have been rather few studies on stability of nonlinear positive systems with time-varying delays. An important reason for this is that popular techniques for analyzing positive systems with constant delays, such as linear Lyapunov-Krasovskii functionals, cannot be applied or lead to excessive conservatism when the delays are time-varying.

At this point, it is worth noting that the results for homogeneous cooperative systems and positive linear systems cited above concern \textit{asymptotic} stability. However, there are processes and applications for which it is desirable that the system converges quickly enough to the equilibrium. While \textit{exponential} stability of positive linear systems with constant delays was investigated in~\cite{Zhu:12} using Lyapunov-Krasovskii techniques, extensions to time-varying delays are non-trivial. Moreover, although quantitative stability measures can be highly dependent on the magnitude of delays, no sharp characterization of how a maximum delay bound affects the guaranteed decay rate of a positive system exists to date. This paper addresses these issues.

At the core of our paper is a set of powerful conditions for establishing exponential stability of a particular class of nonlinear continuous- and discrete-time positive systems with bounded time-varying delays. More specifically, we make the following contributions:
\begin{itemize}
\item[\textbf{1)}] We derive a \emph{necessary and sufficient} condition for exponential stability of continuous-time positive systems whose constituent vector fields are homogeneous of degree one and cooperative. 
\item[\textbf{2)}] For the case which the time delays have a known upper bound, we present an explicit expression that bounds the decay rate of the system. 
\item[\textbf{3)}] We demonstrate that the best decay rate of positive linear systems that our bound can provide can be found via convex optimization techniques. 
\item[\textbf{4)}] We  extend our obtained results to general linear systems with time-varying delays. 
\item[\textbf{5)}] Finally, we provide the corresponding counterparts for discrete-time positive systems.
\end{itemize}

The remainder of the paper is organized as follows. In Section \ref{sec:preliminaries}, we review some required background results and introduce the notation that will be used throughout the paper. The main results of this work for continuous- and discrete-time positive systems are stated in Sections~\ref{sec:Continuous-Time Case} and~\ref{sec:Discrete-Time Case}, respectively. Illustrative examples are presented in Section~\ref{sec:examples}, justifying the validity of our results. Finally, concluding remarks are given in Section~\ref{sec:conclusions}.

%
%

\section{Notation and Preliminaries}\label{sec:preliminaries}

Vectors are written in bold lower case letters and matrices in capital letters. We let $\mathbb{R}$, $\mathbb{N}$, and $\mathbb{N}_0$ denote the set of real numbers, natural numbers, and the set of natural numbers including zero, respectively. The non-negative orthant of the \textit{n}-dimensional real space $\mathbb{R}^n$ is represented by~$\mathbb{R}^n_+$. The $i^{th}$ component of a vector $\x\in \mathbb{R}^n$ is denoted by $x_i$, and the notation $\x\geq \y$ means that $x_i\geq y_i$ for all components~$i$. Given a vector $\v>\bnull$, the weighted $l_\infty$ norm is defined by
\begin{align*}
\|\x\|_\infty^{\v}&=\max_{1\leq i\leq n}{\frac{|x_i|}{v_i}}.
\end{align*}
For a matrix~$A=[a_{ij}]\in \mathbb{R}^{n\times n}$, $a_{ij}$ denotes the entry in row $i$ and column~$j$, and $|A|$ is the matrix whose elements are $|a_{ij}|$. A matrix $A\in \mathbb{R}^{n\times n}$ is said to be \textit{non-negative} if $a_{ij}\geq 0$ for all $i$,~$j$. It is called \textit{Metzler} if $a_{ij}\geq 0$ for all $i\neq j$. For a real interval~$[a,b]$, $\mathcal{C}\bigl([a,b],\mathbb{R}^{n}\bigr)$ denotes the space of all real-valued continuous functions on $[a,b]$ taking values in $\mathbb{R}^{n}$. The upper-right Dini-derivative of a continuous function $h:\mathbb{R}\rightarrow \mathbb{R}$ is denoted by $D^+h(\cdot)$.

Next, we review the key definitions and results necessary for developing the main results of this paper. We start with the definition of~\textit{cooperative} vector fields.
\begin{definition}
A continuous vector field $f:\mathbb{R}^{n} \rightarrow \mathbb{R}^{n}$ which is continuously differentiable on~$\mathbb{R}^{n}\backslash\{\bnull\}$ is said to be cooperative if the Jacobian matrix $\frac{\partial f}{\partial x}(\a)$ is Metzler for all $\a\in\mathbb{R}^{n}_+\backslash\{\bnull\}$.
\end{definition}
The next proposition provides an important property of cooperative vector fields. 
\begin{proposition}\textup{\textbf{\cite[Chapter 3, Remark 1.1]{Smith:95}}}
Let $f:\mathbb{R}^{n} \rightarrow \mathbb{R}^{n}$ be a cooperative vector field. For any two vectors $\x$ and $\y$ in $\mathbb{R}^{n}_+\backslash\{\bnull\}$ with $x_i=y_i$ and $\x\geq \y$, we have $f_i(\x)\geq f_i(\y)$.
\label{Proposition:1}
\end{proposition}
The following definition introduces~\textit{homogeneous} vector fields.
\begin{definition}
A vector field $f:\mathbb{R}^{n} \rightarrow \mathbb{R}^{n}$ is called homogeneous of degree $\alpha$ if for all~$\x \in \mathbb{R}^{n}$ and all real $\lambda>0$, $\f(\lambda\x)=\lambda^{\alpha}\f(\x)$.
\end{definition}
When $\alpha=1$, then $f$ is called the~\textit{homogeneous of degree one}.
Finally, we recall the definition of an \textit{order-preserving} vector field.
\begin{definition}
A vector field $g:\mathbb{R}^{n} \rightarrow \mathbb{R}^{n}$ is said to be order-preserving on $\mathbb{R}^{n}_+$ if $\g(\x)\geq \g(\y)$ for any $\x,\y\in\mathbb{R}^{n}_+$ such that $\x\geq \y$.
\end{definition}

%

\section{Continuous-Time Case}\label{sec:Continuous-Time Case}


Consider the continuous-time nonlinear dynamical system
\begin{align}
{\mathcal G:}
& \left\{
\begin{array}[l]{ll}
\dot{\x}\bigl(t\bigr)=\f\bigl(\x(t)\bigr)+\g\bigl(\x(t-\tau(t))\bigr),& t\geq 0,\\
\x\bigl(t\bigr)=\bphi\bigl(t\bigr), & t\in[-\tau_{\max},0].
\end{array}
\right.
\label{System:1}
\end{align}
Here, $\tau_{\max}\geq 0$, $\x(t)\in\mathbb{R}^{n}$ is the system state, $f,g:~\mathbb{R}^{n} \rightarrow \mathbb{R}^{n}$ are system vector fields with $\f(\bnull)=\g(\bnull)=\bnull$, and $\bphi(\cdot)\in \mathcal{C}\bigl([-\tau_{\max},0],\mathbb{R}^{n}\bigr)$ is the vector-valued initial function specifying the initial state of the system. The delay $\tau(\cdot)$ is assumed to be continuous with respect to time, not necessarily continuously differentiable, and satisfies $0\leq \tau(t)\leq \tau_{\max}$ for all $t\geq 0$. While no restriction on the derivative of $\tau(t)$ (such as $\dot{\tau}<1$) is imposed, causality of the state space for system~\eqref{System:1} even under fast-varying delays is preserved, since~$\tau(\cdot)$ is assumed to be bounded~\cite{Verriest:11}.

In the remainder of the section, vector fields $f$ and $g$ satisfy Assumption~\ref{Assumption:1}.
\begin{assump}\label{Assumption:1}
The following properties hold.
\begin{enumerate}
\item[a)] $f$ and $g$ are continuous on $\mathbb{R}^{n}$, continuously differentiable on $\mathbb{R}^{n}\backslash\{\bnull\}$, and homogeneous;
\item[b)] $f$ is cooperative and $g$ is order-preserving on $\mathbb{R}^{n}_+$.
\end{enumerate}
\end{assump}
Assumption 1a) implies that $f$ and $g$ are globally Lipschitz on $\mathbb{R}^{n}$ \cite[Lemma 2.1] {Mason:09}. Since $\bphi(\cdot)$ and $\tau(\cdot)$ are continuous functions of time, it then follows that there exists a unique $\x(t)$ defined on $[-\tau_{\max},\infty)$ that coincides with $\bphi(\cdot)$ on $[-\tau_{\max},0]$ and satisfies~\eqref{System:1} for $t\geq 0$~\cite[pp. 408--409]{Driver:62}.

The time-delay dynamical system~$\mathcal{G}$ given by~\eqref{System:1} is said to be \textit{positive} if for every non-negative initial condition $\bphi(\cdot)\in \mathcal{C}\bigl([-\tau_{\max},0],\mathbb{R}^{n}_+\bigr)$, the corresponding state trajectory is non-negative, that is $\x(t)\in\mathbb{R}_+^{n}$ for all $t\geq 0$. It follows from~\cite[Chapter 5, Theorem 2.1]{Smith:95} that Assumption~1b) ensures the positivity of system~$\mathcal{G}$ given by~\eqref{System:1}.

While~$\x=\bnull$ is clearly an equilibrium point of the system~\eqref{System:1}, it is not necessarily stable. Moreover, the stability of general systems may depend on the magnitude and variation of the time delays. However, it was shown in~\cite[Theorem 4.1]{Mason:09} that under Assumption~\ref{Assumption:1}, the positive system~\eqref{System:1} with constant delays $(\tau(t)=\tau_{\max}\;\textup{for all}\;t\geq 0)$ is globally asymptotically stable for all $\tau_{\max}\geq 0$ if and only if the undelayed system $(\tau_{\max}= 0)$ is globally asymptotically stable. Our main objectives are therefore $(i)$ to determine if a similar delay-independent stability result holds for the homogeneous cooperative system~\eqref{System:1} with bounded time-varying delays; and $(ii)$ to determine how the decay rate of the positive system~\eqref{System:1} depends on the magnitude of time delays. 

The following theorem establishes a necessary and sufficient condition for exponential stability of homogeneous cooperative systems with bounded time-varying delays and is our first key result. 

\begin{theorem}\label{Theorem:1}
For system~$\mathcal{G}$ given by~\eqref{System:1}, suppose Assumption~\ref{Assumption:1} holds. The following statements are equivalent.
\begin{itemize}
\item[(a)] There exists a vector $\v>\bnull$ such that
\begin{align}
\f(\v)+\g(\v)<\bnull\label{Theorem:1-0}.
\end{align}
\item[(b)] The positive system~$\mathcal{G}$ is globally exponentially stable for all bounded time delays. In particular, every solution $\x(t)$ of~$\mathcal{G}$ satisfies
\begin{align*}
\| \x(t)\|_{\infty}^{\v} \leq  \|\bphi\|e^{-\eta t},\quad t\geq 0,
\end{align*}
where $\|\bphi\|=\sup_{-\tau_{\max}\leq s\leq 0}\|\bphi(s)   \|_{\infty}^{\v}$, $\eta\in\bigl(0,\min_{1\leq i\leq n} \eta_i\bigr)$, and $\eta_i$ is the unique positive solution of the equation
\begin{align}
\left(\frac{f_i(\v)}{v_i}\right)+\left(\frac{g_i(\v)}{v_i}\right)e^{\eta_i\tau_{\max}}+\eta_i=0,\quad i=1,\ldots,n.\label{Theorem:1-1}
\end{align}
\end{itemize}
\end{theorem}

\begin{IEEEproof}
See Appendix~\ref{appendix:1}.
\end{IEEEproof}

\begin{remark}\label{Remark:0}\textup{
Equation~\eqref{Theorem:1-1} has three parameters: the positive vector $\v$, $\tau_{\max}$, and $\eta_i$. For any fixed $\v>\bnull$ and $\tau_{\max}\geq 0$, \eqref{Theorem:1-1} is a nonlinear equation with respect to $\eta_i$. The left-hand side of~\eqref{Theorem:1-1} is strictly monotonically increasing in $\eta_i>0$ and, by~\eqref{Theorem:1-0}, is smaller than the right-hand side for $\eta_i=0$. Therefore,~\eqref{Theorem:1-1} always admits a unique positive solution $\eta_i$.
}
\end{remark}

According to Theorem~\ref{Theorem:1}, the homogeneous cooperative system~$\mathcal{G}$ given by~\eqref{System:1} is globally exponentially stable for all bounded delays if and only if the the corresponding system without delay is stable. In other words, the exponential stability does not depend on the magnitude of the delays, but only on the vector fields. Moreover, any vector $\v>\bnull$ satisfying~\eqref{Theorem:1-0} can be used to find a guaranteed decay rate of the positive system~$\mathcal{G}$ by computing the associated $\eta$. Note that $\eta_i$ in~\eqref{Theorem:1-1} is monotonically decreasing in $\tau_{\max}$ and approaches zero as $\tau_{\max}$ tends to infinity. Hence, the guaranteed decay rate deteriorates with increasing~$\tau_{\max}$.

\begin{remark}\label{Remark:1}\textup{
It has been shown in~\cite[Proposition 3.1]{Mason:09} that~\eqref{Theorem:1-0} has a feasible solution $\v>\bnull$ if and only if there does not exist a non-zero vector~$\w\geq \bnull$ satisfying $\f(\w)+\g(\w)\geq\bnull$. This result provides an alternative test for checking the global exponential stability of the homogeneous cooperative system~$\mathcal{G}$ with time-varying delays.
}
\end{remark}

\begin{remark}\label{Remark:2}\textup{
The result in Theorem~\ref{Theorem:1} can be easily extended to positive nonlinear systems with multiple delays of the form
\begin{align*}
\dot{\x}\bigl(t\bigr)&=\f\bigl(\x(t)\bigr)+\sum _{s=1}^p\g_s\bigl(\x(t-\tau_s(t))\bigr).
\end{align*}
Here,~$p\in\mathbb{N}$, $f:\mathbb{R}^{n} \rightarrow \mathbb{R}^{n}$ is cooperative and homogeneous of degree one, $g_s:\mathbb{R}^{n} \rightarrow \mathbb{R}^{n}$ for~$s=1,\ldots,p$ are homogenous and order-preserving on $\mathbb{R}^{n}_+$, and~$0\leq\tau_s(t)\leq \tau_{\max}$ for $t\geq 0$. In this case, the stability condition~\eqref{Theorem:1-0} becomes
\begin{align*}
\f(\v)+\sum _{s=1}^p \g_s(\v)<\bnull.
\end{align*}
}
\end{remark}

We now discuss delay-independent exponential stability of a special case of~\eqref{System:1}, namely the continuous-time linear dynamical system~$\mathcal{G}_L$ of the form
\begin{align}
{\mathcal G}_{L}:
& \left\{
\begin{array}[l]{ll}
\dot{\x}\bigl(t\bigr)=A\x\bigl(t\bigr)+B\x\bigl(t-\tau(t)\bigr),& t\geq 0,\\
\x\bigl(t\bigr)=\bphi\bigl(t\bigr), & t\in[-\tau_{\max},0].
\end{array}
\right.
\label{System:2}
\end{align}
In terms of (\ref{System:1}), $\f(\x)=A\x$ and $\g(\x)=B\x$. It is easy to verify that if $A$ is Metzler and $B$ is non-negative, Assumption~\ref{Assumption:1} is satisfied. We then have the following special case of Theorem~\ref{Theorem:1}.

\begin{theorem}\label{Theorem:2}
Consider linear system~$\mathcal{G}_L$ given by~\eqref{System:2} where $A$ is Metzler and $B$ is non-negative. Then, there exists a vector $\v > \bnull$ such that
\begin{align}\label{LP}
\bigl(A+B\bigr)\v<\bnull,
\end{align}
if and only if the positive system~$\mathcal{G}_L$ is globally exponentially stable for all bounded delays.
\end{theorem}

The stability condition~\eqref{LP} is a linear programming problem in $\v$, and thus can be verified numerically in polynomial time. Clearly, the exponential bound on the decay rate of positive linear systems that our results can ensure depends on the choice of vector $\v$, and that an arbitrary feasible $\v$ not necessarily gives a tight bound on the actual decay rate. However, we will show that the best guaranteed decay rate can be found via convex optimization. To this end, we use the change of variables $z_i=\textup{ln}(v_i)$, $i=1,\ldots,n$. Then, the search for $\v$ can be formulated as
\begin{subequations}\label{convex}
\begin{align}
&\hspace{-2.0cm} \textbf{maximize}\hspace{0.5cm} \eta\nonumber\\
&\hspace{-2.0cm}\textbf{subject to}\hspace{0.3cm} \eta< \eta_i,\\
&\hspace{0.3cm}a_{ii}+b_{ii}+\sum_{j\neq i} \bigl(a_{ij}+b_{ij}\bigr)e^{z_j-z_i}<0,\\
&\hspace{0.3cm}a_{ii}+\sum_{j\neq i} a_{ij}e^{z_j-z_i}+\sum_{j=1}^n b_{ij}e^{z_j-z_i+\eta_i\tau_{\max}}+\eta_i\leq 0,\quad i=1,\ldots,n,
\end{align}
\end{subequations}
where the last two constraints are~\eqref{LP} and~\eqref{Theorem:1-1} in the new variables, respectively. The optimization variables are the decay rate $\eta$ and the vector $\z=[z_1,\ldots,z_n]^T$. Since $a_{ij}\geq 0$ for all $i\neq j$ and $b_{ij}\geq 0$ for all $i,j$, the last two constraints in~\eqref{convex} are convex in $\eta$ and $z$. This implies that this is a convex optimization problem; hence, it can be efficiently solved.

\begin{remark}\label{Remark:4}\textup{
A necessary and sufficient condition for asymptotic stability of the positive linear system~\eqref{System:2} with time-varying delays has been established in~\cite{Liu:10}. Moreover, in~\cite{Liu:13}, it has been shown that if~\eqref{System:2} is asymptotically stable for all bounded delays, it is also exponentially stable for all bounded delays. However, the impact of delays on the decay rate of~\eqref{System:2} was missing in~\cite{Liu:10,Liu:13}. Thus, not only do we extend the result of~\cite{Liu:13} to general homogeneous cooperative systems (not necessarily linear), but we also provide an explicit exponential bound on the decay rate.
}
\end{remark}

We now extend Theorem~\ref{Theorem:2} to general linear systems, not necessarily positive.

\begin{theorem}\label{Theorem:3}
Suppose that there exists a vector $\v>\bnull$ such that
\begin{align}\label{LP-1}
\bigl(A^M+|B|\bigr)\v<\bnull,
\end{align}
where $A^M=[a_{ij}^M]$ is a matrix with $a^M_{ii}=a_{ii}$ and $a^M_{ij}=|a_{ij}|$ for all $i\neq j$. Let $\eta_i$ be the unique positive solution of the equation
\begin{eqnarray}
\biggl(a_{ii}+\sum_{j\neq i} \frac{1}{v_i}|a_{ij}|v_j\biggr)+\biggl(\sum_{j=1}^n \frac{1}{v_i}\bigl|b_{ij}|v_j\biggr)e^{\eta_i\tau_{\max}}+\eta_i=0.\label{Theorem:2-3-1}
\end{eqnarray}
Then, linear system~$\mathcal{G}_L$ given by~\eqref{System:2} is globally exponentially stable. Furthermore,
\begin{align*}
\| \x(t)\|_{\infty}^{\v} \leq   \|\bphi\|e^{-\eta t}, \ \ t\geq 0,
\end{align*}
where $0<\eta<\min_{1\leq i\leq n} \eta_i$.
\end{theorem}

\begin{IEEEproof}
See Appendix~\ref{appendix:2}.
\end{IEEEproof}

\begin{remark}\label{Remark:5}\textup{
The stability condition~\eqref{LP-1} does not include any information on the magnitude of delays, so it ensures \textit{delay-independent} stability. Since $A^M$ is Metzler and $|B|$ is non-negative, $A^M+|B|$ is Metzler. It follows from~\cite[Proposition 2]{Rantzer:11} that inequality~\eqref{LP-1} holds if and only if $A^M+|B|$ is Hurwitz, \textit{i.e.}, all its eigenvalues have negative real parts.
}
\end{remark}

\section{Discrete-time Case}\label{sec:Discrete-Time Case}

Next, we consider the discrete-time analog of (\ref{System:1}):
\begin{align}
{\Sigma}:
& \left\{
\begin{array}[l]{ll}
\x\bigl(k+1\bigr)=\f\bigl(\x(k)\bigr)+\g\bigl(\x(k-d(k))\bigr),& k \in \mathbb{N}_0\\
\x\bigl(k\bigr)=\bph\bigl(k\bigr), &k\in\{-d_{\max},\ldots,0\}.
\end{array}
\right.
\label{System:3}
\end{align}
Here, $\x(k)\in\mathbb{R}^{n}$ is the state variable, $f,g:~\mathbb{R}^{n} \rightarrow \mathbb{R}^{n}$, $\f(\bnull)=\g(\bnull)=\bnull$, $d_{\max}\in\mathbb{N}_0$, $d(k)\in\mathbb{N}_0$ represents the time-varying delay satisfying $0\leq d(k)\leq d_{\max}$ for all $k\in\mathbb{N}_0$, and $\bph(\cdot):\{-d_{\max},\ldots,0\} \rightarrow \mathbb{R}^{n}$ is the vector sequence specifying the initial state of the system. For the remainder of this section, Assumption~\ref{Assumption:4} holds.
\begin{assump}\label{Assumption:4}
$f$ and $g$ are continuous on $\mathbb{R}^{n}$, homogeneous of degree one, and order-preserving on $\mathbb{R}^{n}_+$.
\end{assump}

The time-delay dynamical System~$\Sigma$ given by~\eqref{System:3} is said to be \textit{positive} if for every non-negative initial condition $\bph(\cdot)\in\mathbb{R}^{n}_+$, the corresponding solution is non-negative, \textit{i.e.}, $\x(k)\geq \bnull$ for all $k \in \mathbb{N}$. Note that under Assumption~\ref{Assumption:4}, system~$\Sigma$ is positive.

Next theorem shows that homogeneous monotone systems are insensitive to bounded delays.

\begin{theorem}\label{Theorem:4}
For system~$\Sigma$ given by~\eqref{System:3}, suppose Assumption~\ref{Assumption:4} holds. Then, the following statements are equivalent.
\begin{itemize}
\item[(a)] There exists a vector $\v>\bnull$ such that
\begin{align}
\f(\v)+\g(\v)<\v\label{Theorem:3-0}.
\end{align}
\item[(b)] The positive system~$\Sigma$ is globally exponentially stable for all bounded time delays. In particular, every solution $\x(k)$ of~$\Sigma$ satisfies
\begin{align}
\| \x(k)\|_{\infty}^{\v} \leq  \|\bph\|\gamma^{k},\quad k\in \mathbb{N}_0,\label{Theorem:3-1}
\end{align}
where $\|\bph\|=\sup_{-d_{\max} \leq s\leq 0}\|\bph(s)   \|_{\infty}^{\v}$, $\gamma=\max_{1\leq i\leq n} \gamma_i$, and $\gamma_i\in(0,1)$ is the unique positive solution of the equation
\begin{align}
\left(\frac{f_i(\v)}{v_i}\right)+\left(\frac{g_i(\v)}{v_i}\right)\gamma_i^{-d_{\max}}=\gamma_i.\label{Theorem:3-2}
\end{align}
\end{itemize}
\end{theorem}

\begin{IEEEproof}
See Appendix~\ref{appendix:3}.
\end{IEEEproof}
Theorem~\ref{Theorem:4} provides a test for the global exponential stability of the homogeneous monotone system~\eqref{System:3} with time-varying delays. In addition, for any vector $\v>\bnull$ that satisfies~\eqref{Theorem:3-0}, this theorem provides an explicit bound on the impact that an increasing delay has on the decay rate. Note that $\gamma_i$ is monotonically increasing in $d_{\max}$, and approaches one as $d_{\max}$ tends to infinity. Hence, the guaranteed decay rate slows down as the delays increase in magnitude.

Let $\f(\x)=A\x$ and $\g(\x)=B\x$ such that $A,B\in\mathbb{R}^{n\times n}$ are non-negative matrices. Then, homogeneous monotone system~\eqref{System:3} reduces to the positive linear system~$\Sigma_L$ of the form
\begin{align}
{\Sigma}_L:
& \left\{
\begin{array}[l]{ll}
{\x}\bigl(k+1\bigr)=A\x\bigl(k\bigr)+B\x\bigl(k-d(k)\bigr), & k \in \mathbb{N}_0\\
\x\bigl(k\bigr)=\bph\bigl(k\bigr), &k\in\{-d_{\max},\ldots,0\}.
\end{array}
\right.
\label{System:4}
\end{align}
Theorem~\ref{Theorem:4} helps us to derive a necessary and sufficient condition for exponential stability of discrete-time positive linear systems. Specifically, we note the following.
\begin{theorem}\label{Theorem:5}
Consider linear system~$\Sigma_L$ given by~\eqref{System:4} where $A$ and $B$ are non-negative. Then, there exists a vector $\v > \bnull$ such that
\begin{align}\label{LP-d}
\bigl(A+B\bigr)\v<\v,
\end{align}
if and only if the positive system~$\Sigma_L$ is globally exponentially stable for all bounded delays.
\end{theorem}

In order to find the best decay rate of the positive linear system~\eqref{System:4} that our bound can provide, we use the logarithmic change of variables $z_i=\textup{ln}(v_i)$ and $\bar{\gamma}_i=\textup{ln}(\gamma_i)$. Note that these change of variables are valid since the variables $v_i$ and $\gamma_i$ are required to be positive for all $i$. Then, the search for vector $\v$ can be formulated as
\begin{subequations}\label{GP}
\begin{align}
&\hspace{-2.0cm} \textbf{minimize}\hspace{0.5cm} e^{\bar{\gamma}}\nonumber\\
&\hspace{-2.0cm}\textbf{subject to}\hspace{0.3cm} e^{\bar{\gamma}_i-\bar{\gamma}}\leq 1,\\
&\hspace{0.3cm}\sum_{j=1}^n \bigl(a_{ij}+b_{ij}\bigr)e^{z_j-z_i}<1,\\[-0.2cm]
&\hspace{0.3cm}\sum_{j=1}^n a_{ij}e^{z_j-z_i-\bar{\gamma_i}}+\sum_{j=1}^n b_{ij}e^{z_j-z_i-\bar{\gamma_i}(d_{\max}+1)}\leq 1,\quad i=1,\ldots,n,
\end{align}
\end{subequations}
where the last two constraints are~\eqref{LP-d} and~\eqref{Theorem:3-2} in the new variables, respectively. Here, the optimization variables are the vector $\z=[z_1,\ldots,z_n]^T$ and $\bar{\gamma}$. Since the constraints in~\eqref{GP} define a convex set and the objective function is convex,~\eqref{GP} is a convex optimization problem. This implies that it can be solved globally and efficiently.

We now give an extension of Theorem~\ref{Theorem:5} to general linear systems with time-varying delays.

\begin{theorem}\label{Theorem:6}
Suppose that there exists a vector $\v>\bnull$ such that
\begin{align}
\bigl(|A|+|B|\bigr)\v<\v.\label{Theorem:6-0}
\end{align}
Let $\gamma_i$ be the positive solution of the equation
\begin{align}
\left(\sum_{j=1}^n \frac{1}{v_i}\bigl|a_{ij}|v_j\right)+\biggl(\sum_{j=1}^n \frac{1}{v_i}\bigl|b_{ij}|v_j\biggr)\gamma_i^{-d_{\max}}=\gamma_i.\label{Theorem:6-1}
\end{align}
Then, the discrete-time linear system~\eqref{System:4} is globally exponentially stable. Moreover,
\begin{align}
\| \x(k)\|_{\infty}^{\v} \leq  \gamma^{k}\|\bph\|,\quad k \in \mathbb{N}_0,\label{Theorem:6-2}
\end{align}
where $\gamma=\max_{1\leq i \leq n} \gamma_i$.
\end{theorem}

\begin{IEEEproof}
See Appendix~\ref{appendix:4}.
\end{IEEEproof}

\section{Illustrative Examples}\label{sec:examples}

\subsection{Continuous-time Nonlinear Positive System}
Consider continuous-time nonlinear dynamical system~$\mathcal{G}$ given by~\eqref{System:1} with
\begin{align}
\f(x_1,x_2)&=\begin{bmatrix} -3 & 6\\ 2 & -2 \end{bmatrix}\begin{bmatrix} x_1 \\ x_2 \end{bmatrix}-\sqrt{x_1^2+x_2^2}\begin{bmatrix} 3 \\ 1\end{bmatrix},\quad \g(x_1,x_2)=\begin{bmatrix}\frac{x_1x_2}{\sqrt{x_1^2+x_2^2}}\\ \frac{x_1x_2}{\sqrt{2x_1^2+3x_2^2}} \end{bmatrix}.\label{Example:1}
\end{align}
It is straightforward to verify that both $f$ and $g$ satisfy Assumption~\ref{Assumption:1}~\cite[Example 4.1]{Mason:09}. Moreover, $\f(1,1)+\g(1,1)<\bnull$. It follows from Theorem~\ref{Theorem:1} that~\eqref{Example:1} is globally exponentially stable for all bounded time delays. For example, let $\tau(t)= 5+\sin(t)$ and set $\tau_{\max} = 6$. By using the vector $\v=~(1 , 1)$ together with $\tau_{\max}=6$, the solutions to the equation~\eqref{Theorem:1-1} can be obtained as $\eta_1=0.0825$ and $\eta_2=0.1705$. Thus, the decay rate of positive system~\eqref{Example:1} is upper bounded by $\eta\approx\min\{0.0825,0.1705\}=0.0825$. In particular, $
\| \x(t)\|_{\infty}^{\v} \leq  \|\bphi\|e^{-0.0825 t}$ for all $t\geq 0$. Figure~\ref{Fig:2} gives the simulation results of the actual decay rate of positive system~\eqref{Example:1}, $x_1(t)$ and $x_2(t)$, and the theoretical upper bound $e^{-0.0825 t}$ when the initial condition is $\bphi(t) = (1,1)$ for $t\in[-6,0]$. Note that~\cite[Theorem 4.1]{Mason:09} can not be applied in this example to ascertain the stability of homogeneous cooperative system~\eqref{Example:1}, since the delay is assumed to be time-varying.
\begin{figure}[h]
\centering
\includegraphics [width=0.48\columnwidth]{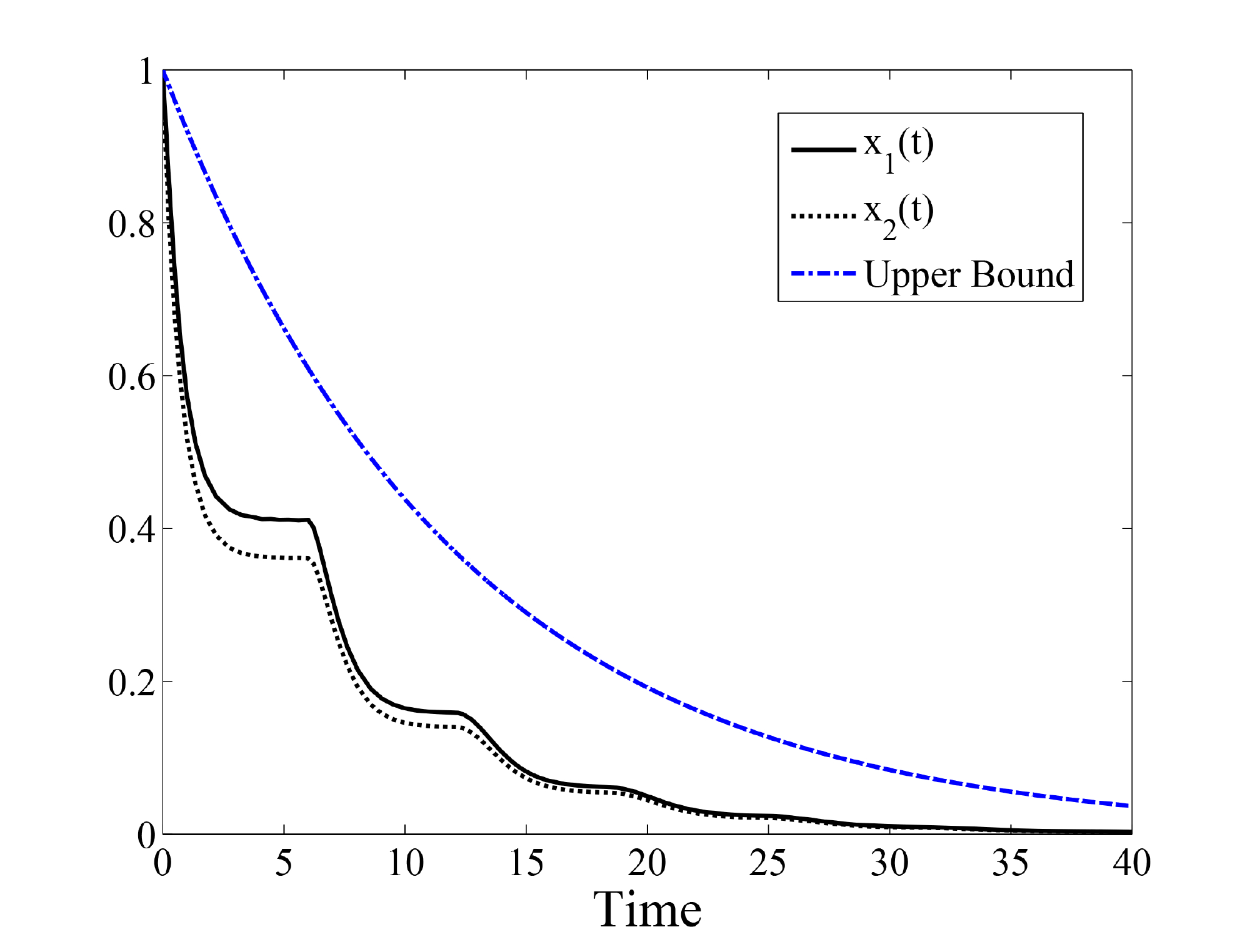}
\caption{Comparison of upper bound and actual decay rate of positive system~\eqref{Example:1} with bounded time-varying delays.} \label{Fig:2}
\end{figure}

\subsection{Continuous-time Linear Positive System}

\begin{figure}[ht]
\begin{minipage}[b]{0.48\linewidth}
\centering
\includegraphics[width=\textwidth]{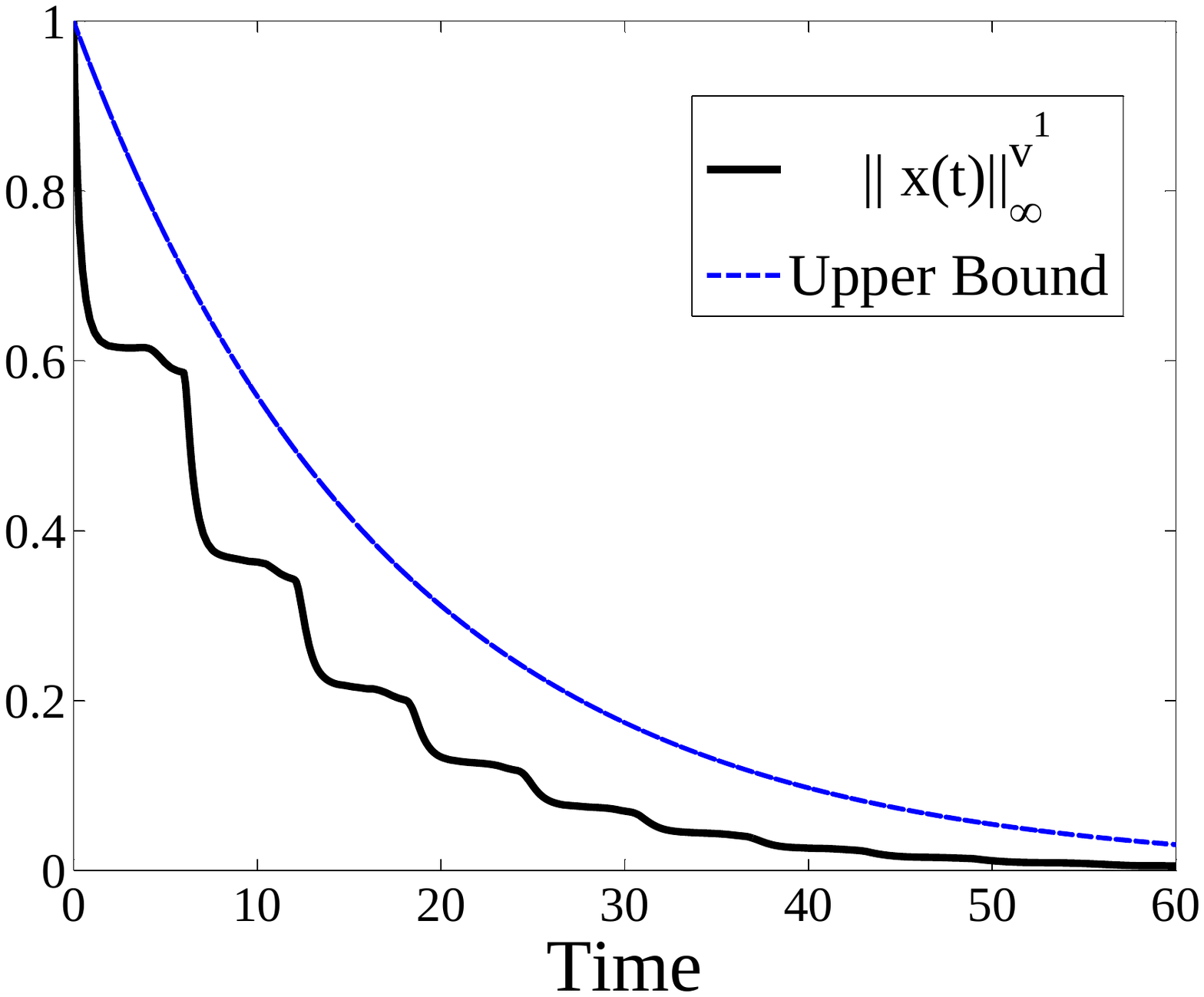}
\end{minipage}
\hspace{0.5cm}
\begin{minipage}[b]{0.48\linewidth}
\centering
\includegraphics[width=\textwidth]{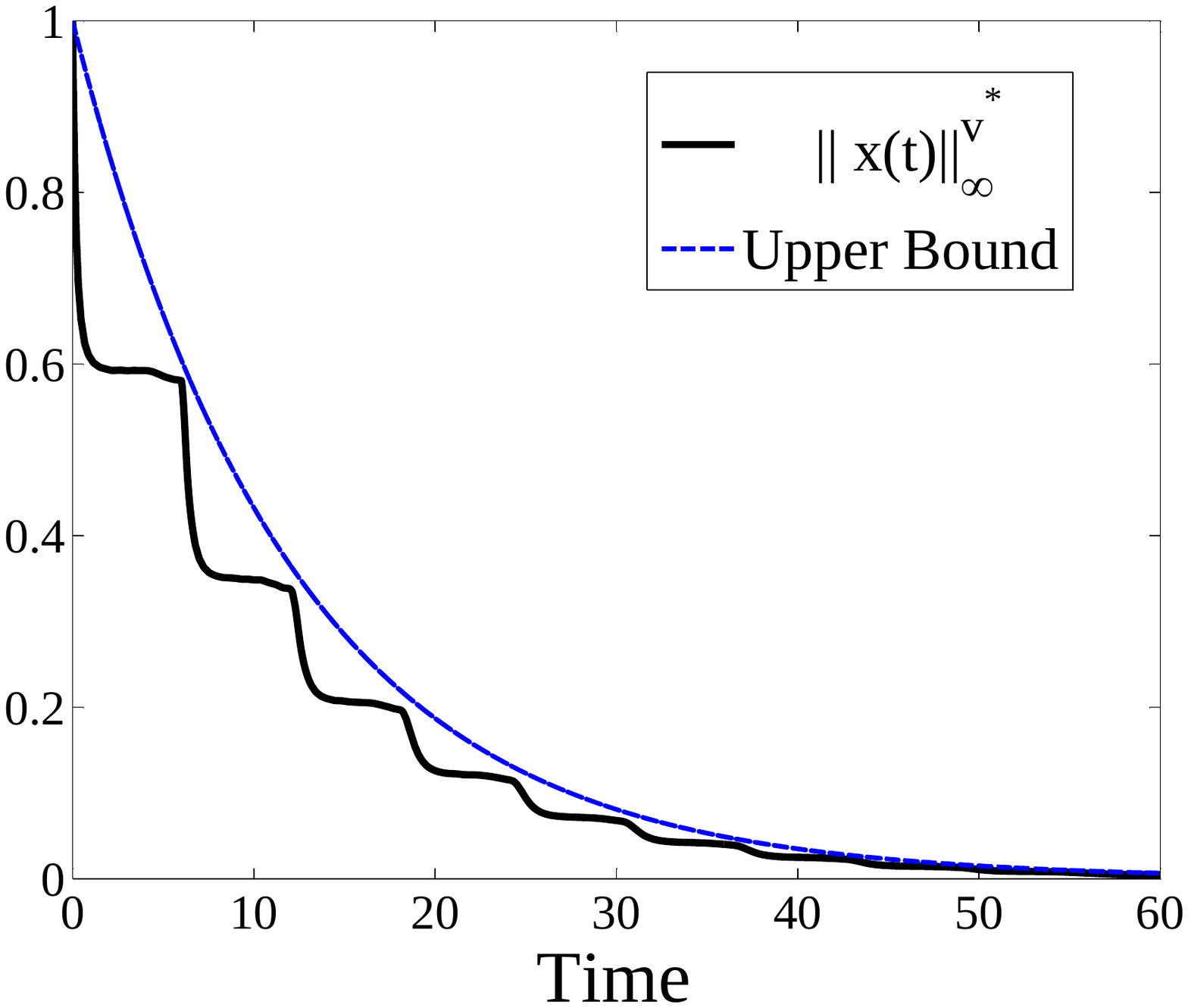}
\end{minipage}
\caption{Comparison of upper bounds and actual decay rates of the solution $\x(t)$ without (left) and with (right) convex optimization  for the system described by~\eqref{Example:2}.} \label{Fig:1}
\end{figure}

\noindent Consider the continuous-time linear system~\eqref{System:2} with
\begin{align}
A&=\begin{bmatrix} -6 & 2 \\ 1 & -3\end{bmatrix},\; B=\begin{bmatrix} 3 & 0 \\ 0 & 0.5\end{bmatrix}.
\label{Example:2}
\end{align}
The time-varying delay is given by
\begin{align*}
\tau(t)=5+\textup{sin}(t).
\end{align*}
Obviously, one may choose $\tau=6$ as an upper bound on the delay. Since $A$ is Metzler and $B$ is non-negative, the system~\eqref{Example:2} is \emph{positive}.\\
By Theorem~\ref{Theorem:2}, since $A+B$ is Hurwitz,~\eqref{Example:2} is exponentially stable for any bounded time-varying delays. Moreover, according to the linear inequality~\eqref{LP}, the following inequality must be fulfilled
\begin{align}\label{Example:2-1}
\begin{cases}
\begin{bmatrix} -3 & 2\\ 1 & -2.5\end{bmatrix}\begin{bmatrix}v_1\\ v_2\end{bmatrix}<\bnull,\\
\hspace{0.5cm} v_1,v_2>0.
\end{cases}
\end{align}
As discussed in Section~\ref{sec:Continuous-Time Case}, any feasible solution  $\v$ to these inequalities can be used to find a guaranteed rate of convergence of the system~\eqref{Example:2} by computing the associated $\eta$ in~\eqref{Theorem:1-1}.\\
One natural candidate for $\v$ can be found by considering the delay-free case. The solution of the positive system~\eqref{Example:2} with zero delay, $\dot{\x}(t)=(A+B)\x(t)$, satisfies
\begin{align*}
\| \x(t)\|_{\infty}^{\v} \leq  \|\x(0) \|_{\infty}^{\v}\;e^{\mu_\infty^{\v}(A+B) t},\quad t\geq 0.
\end{align*}
For any vector $\v>\bnull$, since $A+B$ is Metlzer, $\pi(A+B) \leq \mu_\infty^{\v}(A+B)$.
According to the Perron-Frobenius theorem for Metzler matrices~\cite[Theorem 17]{Farina:00}, if $A+B$ is Metzler and \textit{irreducible}, then there exists an eigenvector $\w>\bnull$ such that
\begin{align*}
(A+B)\w&=\pi(A+B)\w.
\end{align*}
It is clear that the vector $\w$ satisfies $\pi(A+B)=\mu_\infty^{\w}(A+B)$.\\
According to the above discussion, one natural candidate $\v$ can be the eigenvector of $A+B$ corresponding to $\pi(A+B)$ which gives the fast decay rate of solutions for the undelayed case. For the system~\eqref{Example:2},
$\pi(A+B)=-1.3139$, and the corresponding eigenvector is
$$
\v^1=\begin{bmatrix}0.7645 & 0.6446\end{bmatrix}^T.
$$
By using this solution together with $\tau=6$, the solutions to the nonlinear equation~\eqref{Theorem:1-1} can be obtained as
$$
\eta_1=0.0583,\;\eta_2=0.1957.
$$
Thus,~\eqref{Example:2} is globally exponentially stable with decay rate $\eta=\min\{0.0583, 0.1957\}=0.0583$. In particular,
\begin{align*}
\| \x(t)\|_{\infty}^{\v^1} \leq   \sup_{-\tau\leq s\leq 0}\left\{\|\bphi(s) \|_{\infty}^{\v^1}\right\}\;e^{-0.0583 t}, \ \ t\geq 0.
\end{align*}
The left-hand side of Figure~\ref{Fig:1} compares $\| \x(t)\|_{\infty}^{\v^{1}}$ obtained by simulating~\eqref{Example:2} from initial condition $\bphi(t)=\v^{1}$ and the theoretical decay rate bound $e^{-0.0583 t}$. Of course, $\v^1$ is only one of the possible solutions of~\eqref{Example:2-1}. Next, by solving the convex optimization problem~\eqref{convex}, we get
$$
\v^{\star}=[0.9020,\; 0.4317]^T,\;\eta^{\star}=0.0838,
$$
which implies that the system~\eqref{Example:2} is globally exponentially stable with decay rate $0.0838$, and the solution $\x(t)$ satisfies
\begin{align*}
\| \x(t)\|_{\infty}^{\v^{\star}} \leq   \sup_{-\tau\leq s\leq 0}\left\{\|\bphi(s) \|_{\infty}^{\v^{\star}}\right\}\;e^{-0.0838 t}.
\end{align*}
The right-hand side of Figure~\ref{Fig:1} gives the simulation results of $\| \x(t)\|_{\infty}^{\v^{\star}}$, and the theoretical upper bound $e^{-0.0838 t}$ when the initial condition is $\bphi(t)=\v^{\star}$.\\
We can see that the linear inequalities~\eqref{Example:2-1} do not help us in guiding our search for a vector $\v$ which guarantees a fast decay rate. In contrast, solving the convex optimization problem~\eqref{convex} finds the best $\eta^{\star}$ that our bound can guarantee along with the associated $\v^{\star}$. The bound matches simulations very well and is a significant improvement over simply using the non-optimized $\v^1$.\\

\subsection{Discrete-time Linear Positive System}

\noindent Consider the discrete-time linear system~\eqref{System:4} with
\begin{align}
A&=\begin{bmatrix} 0.4 & 0.1 \\ 0.2 & 0.6\end{bmatrix},\; B=\begin{bmatrix} 0.3 & 0 \\ 0 & 0.1\end{bmatrix}.
\label{Example:3}
\end{align}
The time-varying delay is given by
$$
d(k)=4+\textup{sin}\left(\frac{k\pi}{2}\right),
$$
with an upper bound~$d=5$. Since $A$ and $B$ are non-negative, the system~\eqref{Example:3} is \emph{positive}.\\
Since $\rho(A+B)<1$, Theorem~\ref{Theorem:5} guarantees that the system~\eqref{Example:3} is exponentially stable and that the following set of inequalities have a solution
\begin{align}\label{Example:3-1}
\begin{cases}
\begin{bmatrix} -0.3 & 0.1\\ 0.2 & -0.3\end{bmatrix}\begin{bmatrix}v_1\\ v_2\end{bmatrix}<\bnull,\\
\hspace{0.5cm} v_1,v_2>0.
\end{cases}
\end{align}
As in the continuous-time example, any feasible solution $\v$ of~\eqref{Example:3-1} yields a guaranteed decay rate of the system~\eqref{Example:3} by computing the associated $\gamma$ in~\eqref{Theorem:3-2}. To find the optimal $\v$ for our bound, we solve the convex optimization problem~\eqref{GP}, to find the vector $\v^{\star}$ and its guaranteed decay rate $\gamma^{\star}$:
$$
\v^{\star}=[0.6884,\; 0.7254]^T,\;\gamma^{\star}=0.9320.
$$
Therefore, the solution $\x(k)$  satisfies
\begin{align*}
\| \x(k)\|_{\infty}^{\v^{\star}} \leq   (0.9320)^k\|\bph\|,\quad k \in \mathbb{N}.
\end{align*}
Figure~\ref{Fig:2} shows a comparison of $\|\x(k)\|_{\infty}^{\v^{\star}}$ and the theoretical bound $(0.9320)^k$, when the initial condition is $\bph(k)=\v^{\star}$.

\begin{figure}[h]
\centering
\includegraphics [width=0.55\columnwidth]{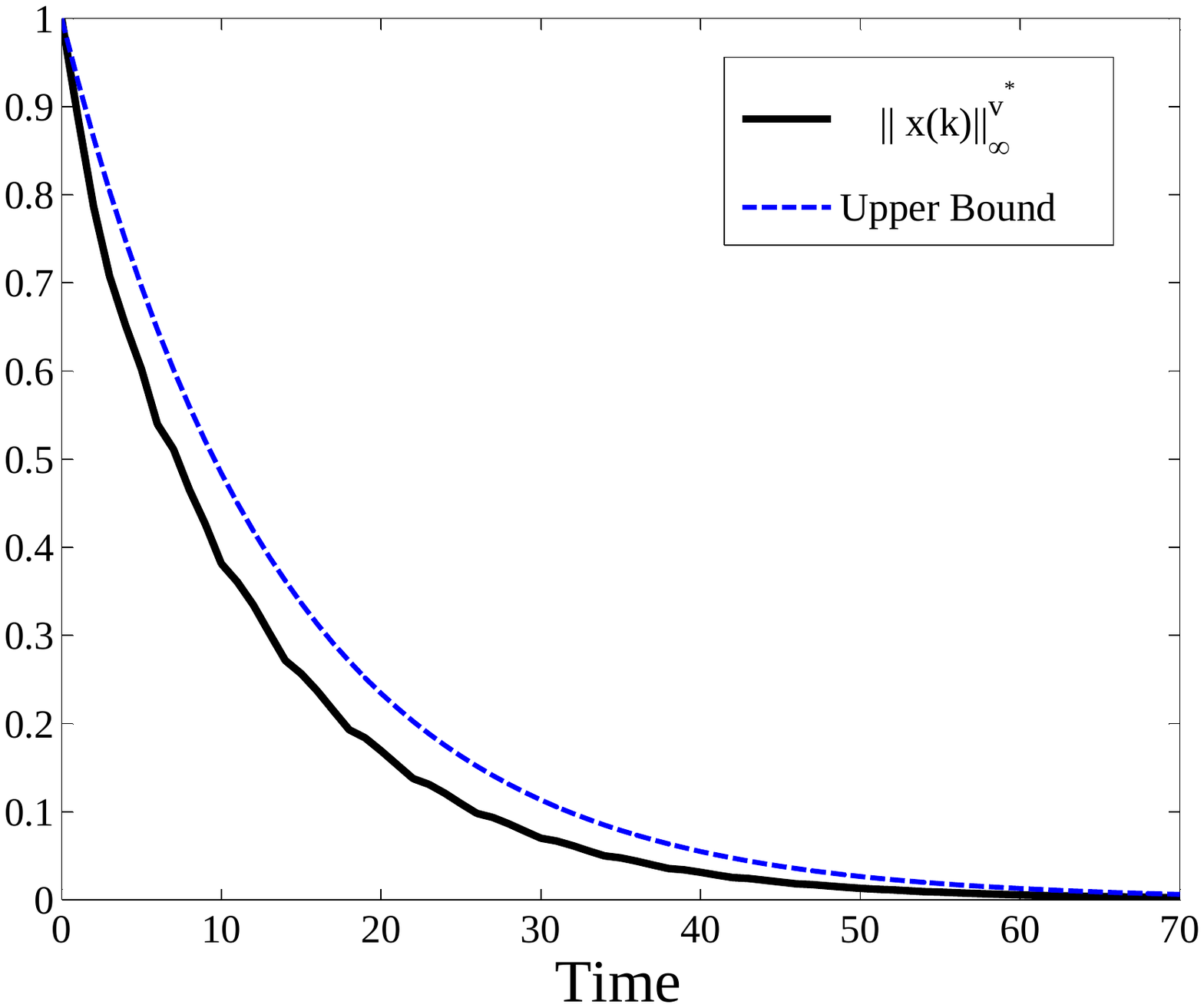}
\caption{Comparison of the upper bound and the actual decay rate of the solution $\x(k)$ for the discrete-time system described by~\eqref{Example:3}.} \label{Fig:2}
\end{figure}

\section{Conclusions}\label{sec:conclusions}

In this paper, we have extended a fundamental property of positive linear systems to a class of nonlinear positive systems. Specifically, we have demonstrated that continuous-time homogeneous cooperative systems and discrete-time homogeneous monotone systems are insensitive to bounded time-varying delays. We have derived a set of necessary and sufficient conditions for establishing delay-independent exponential stability of such positive systems. When the time delays have a known upper bound, explicit expressions that bound the decay rate have been presented. We have further shown that the best bound on the decay rate of positive linear systems that our results can guarantee can be found via convex optimization.  Finally, we have extended obtained results to general linear systems with time-varying delays.

\appendix

\subsection{Proof of Theorem~\ref{Theorem:1}}\label{appendix:1}

$(a)\Rightarrow (b):$ Suppose that there exists a vector~$\v>\bnull$ such that~\eqref{Theorem:1-0} holds. According to Remark~\ref{Remark:0}, Equation~\eqref{Theorem:1-1} always admits a unique positive solution~$\eta_i$. Pick a constant~$\eta$ satisfying $0<\eta<\min_{1\leq i \leq n} \eta_{i}$. Since the left-hand side of~\eqref{Theorem:1-1} is strictly monotonically increasing in $\eta_i>0$, we have
\begin{align}
\left(\frac{f_i(\v)}{v_i}\right)+\left(\frac{g_i(\v)}{v_i}\right)e^{\eta\tau_{\max}}+\eta<0,\quad \textup{for all}\; i.\label{Proof 1-0}
\end{align}
Under Assumption~\ref{Assumption:1}, system~\eqref{System:1} is positive. Hence, $x_i(t)\geq 0$ for all $i$ and all $t\geq 0$. Let
\begin{align}
z_i(t)=\frac{x_i(t)}{v_i}-\|\bphi\|e^{-\eta t}.\label{Proof 1-1}
\end{align}
From the definition of $\|\bphi\|$, $z_i(0)\leq 0$ for all $i$. To prove the exponential stability, we will show that $z_i(t)\leq 0$ for all $i$ and all $t\geq 0$. By contradiction, suppose this is not true. Then, there exist an index $m\in\{1,\ldots,n\}$ and $t_1\geq 0$ such that $z_i(t)\leq 0$ for $t\in[0,t_1]$, $z_m(t_1)=0$, and
\begin{align}
D^+z_m(t_1)&\geq 0.\label{Proof 1-2}
\end{align}
From~\eqref{Proof 1-1}, we have $x_m(t_1)=\|\bphi\|e^{-\eta t_1}v_m$, and $\x(t_1)\leq \|\bphi\|e^{-\eta t_1}\v$. Now, as $f$ is cooperative and homogeneous of degree one,  it follows from Proposition~\ref{Proposition:1} and the above observations that
\begin{align}
f_m\bigl(\x(t_1)\bigr)&\leq f_m\bigl(\|\bphi\|e^{-\eta t_1}\v\bigr)= \|\bphi\|e^{-\eta t_1}f_m\bigl(\v\bigr).\label{Proof 1-4}
\end{align}

\textit{Case 1)} If $\tau(t_1)\leq t_1$, then $t_1-\tau(t_1)\in[0,t_1]$, and therefore $z_i\bigl(t_1-\tau(t_1)\bigr)\leq 0$. As a result,
\begin{align*}
x_i\bigl(t_1-\tau(t_1)\bigr)&\leq   \|\bphi\|e^{-\eta (t_1-\tau(t_1))}v_i\\
&\leq\|\bphi\|e^{-\eta (t_1-\tau_{\max})}v_i,\quad i=1,\ldots,n,
\end{align*}
where we used the fact that $\tau(t_1)\leq \tau_{\max}$ to get the second inequality. Further, as $g$ is order-preserving and homogeneous of degree one, this in turn implies
\begin{align}
g_m\bigl(\x(t_1-\tau(t_1))\bigr)&\leq g_m\bigl(\|\bphi\|e^{-\eta (t_1-\tau_{\max})}\v\bigr)= \|\bphi\|e^{-\eta (t_1-\tau_{\max})}g_m\bigl(\v\bigr).\label{Proof 1-5}
\end{align}
The upper-right Dini-derivative of $z_m(t)$ along the trajectories of~\eqref{System:1} at $t=t_1$ is given by
\begin{align*}
D^+z_m(t_1)&=\frac{\dot{x}_m(t_1)}{v_m}+\|\bphi\|e^{-\eta t_1}\eta\\
&=\frac{f_m\bigl(\x(t_1)\bigr)+g_m\bigl(\x(t_1-\tau(t_1))\bigr)}{v_m}+\|\bphi\|e^{-\eta t_1}\eta\\
&\leq \|\bphi\|e^{-\eta t_1}\left(\left(\frac{f_m(\v)}{v_m}\right)+\left(\frac{g_m(\v)}{v_m}\right)e^{\eta\tau_{\max}}+\eta\right),
\end{align*}
where we substituted~\eqref{Proof 1-4} and~\eqref{Proof 1-5} into the second equality. It now follows from~\eqref{Proof 1-0} that $D^+z_m(t_1)<0$.

\textit{Case 2)} If $\tau(t_1)>t_1$, from the definition of $\|\bphi\|$, we have
$\| \x(t_1-\tau(t_1))\|_{\infty}^{\v} \leq   \|\bphi\|$. Thus, $\x(t_1-\tau(t_1))\leq \|\bphi\|\v$, which implies that
$g_m\bigl(\x(t_1-\tau(t_1))\bigr)\leq \|\bphi\|g_m\bigl(\v\bigr)$. Then,
\begin{align*}
D^+z_m(t_1)&\leq \|\bphi\|e^{-\eta t_1}\left(\left(\frac{f_m(\v)}{v_m}\right)+\left(\frac{g_m(\v)}{v_m}\right)e^{\eta t_1}+\eta\right)\\
&\leq \|\bphi\|e^{-\eta t_1}\left(\left(\frac{f_m(\v)}{v_m}\right)+\left(\frac{g_m(\v)}{v_m}\right)e^{\eta\tau_{\max}}+\eta\right)\\
&<0,
\end{align*}
where the second inequality follows from the fact that $t_1<\tau(t_1)\leq\tau_{\max}$.\\
In summary, we conclude that~$D^+z_m(t_1)<0$, which contradicts~\eqref{Proof 1-2}. Therefore, $z_i(t)\leq 0$ for all $t\geq 0$, and hence $\| \x(t)\|_{\infty}^{\v} \leq   \|\bphi\|e^{-\eta t}$ for $t\geq 0$. This completes the proof.\\
$(b)\Rightarrow (a):$ Assume that system~\eqref{System:1} is exponentially stable for all bounded time delays. Particularly, let $\tau(t)= 0$. Then, $\dot{\x}\bigl(t\bigr)=\f\bigl(\x(t)\bigr)+\g\bigl(\x(t)\bigr)$ is exponentially stable, and hence is asymptotically stable. Since $f+g$ is cooperative and homogeneous of degree one, it follows from~\cite[Theorem 3.1]{Mason:09} that there is some vector~$\v>\bnull$ satisfying~\eqref{Theorem:1-0}.

\subsection{Proof of Theorem~\ref{Theorem:3}}\label{appendix:2}

The proof is almost the same as that of Theorem~\ref{Theorem:1}. From~\eqref{LP-1}, Equation~\eqref{Theorem:2-3-1} always has a unique positive solution~$\eta_i$ for each $i$. Moreover, if $\eta\in\bigl(0,\min_{1\leq i\leq n}\eta_i\bigr)$, then
\begin{align*}
\biggl(a_{ii}+\sum_{j\neq i} \frac{1}{v_i}|a_{ij}|v_j\biggr)+\biggl(\sum_{j=1}^n \frac{1}{v_i}\bigl|b_{ij}|v_j\biggr)e^{\eta\tau_{\max}}+\eta<0,
\end{align*}
hold for all $i$. Let $z_i(t)=|x_i(t)|\slash{v_i}-\|\bphi\|e^{-\eta t}$. We claim that $z_i(t)\leq 0$ for all $t\geq 0$. For each~$i$, the upper-right Dini-derivative of $z_i(t)$ along the trajectories of~\eqref{System:2} is given by
\begin{align*}
D^+z_i(t)&=\frac{\textup{sign}(x_i)}{v_i}\biggl\{\sum_{j=1}^n a_{ij} x_j\bigl(t\bigr)+ \sum_{j=1}^n b_{ij} x_j\bigl(t-\tau(t)\bigr)\biggr\}+\|\bphi\|e^{-\eta t}\eta\nonumber\\
&=\frac{1}{v_i}\biggl\{a_{ii}|x_i(t)|+\textup{sign}(x_i)\sum_{j\neq i} a_{ij} x_j(t)+\textup{sign}(x_i)\sum_{j=1}^nb_{ij} x_j\bigl(t-\tau(t)\bigr)\biggr\}+\|\bphi\|e^{-\eta t}\eta\nonumber\\
&\leq \frac{1}{v_i}\biggl\{a_{ii}\bigl|x_i(t)\bigr|+\sum_{j\neq i} \bigl|a_{ij}\bigr| \bigl|x_j(t)\bigr|+\sum_{j=1}^n \bigl|b_{ij}\bigr| \bigl|x_j(t-\tau(t))\bigr|\biggr\}+\|\bphi\|e^{-\eta t}\eta.
\end{align*}
If there exists an index $m$ and $t_1\geq 0$ such that $z_i(t)\leq 0$ for $t\in[0,t_1]$ and $z_m(t_1)=0$, then the same arguments as in the proof of Theorem~\ref{Theorem:1} yields $D^+z_m(t_1)<0$. The proof is complete.

\subsection{Proof of Theorem~\ref{Theorem:4}}\label{appendix:3}

$(a)\Rightarrow (b):$ First note that, for any fixed $d_{\max}\geq 0$ and any fixed $\v>\bnull$, Equation~\eqref{Theorem:3-2} always has a unique solution $\gamma_i\in(0,1)$~\cite[pp. 444]{BeT:89}. Let $\gamma=\max_{1\leq i\leq n}\gamma_i$. Since the left-hand side of~\eqref{Theorem:3-2} is strictly monotonically decreasing in $\gamma_i$, we have
\begin{align}
\left(\frac{f_i(\v)}{v_i}\right)+\left(\frac{g_i(\v)}{v_i}\right)\gamma^{-d_{\max}}&\leq \gamma_i\leq\gamma,\label{Theorem:5-0}
\end{align}
for all $i$. We now use perfect induction to show that the desired relation~\eqref{Theorem:3-1} is true for all $k\in \mathbb{N}_0$. By the definition of $\|\bph\|$, we have $\|\x(0)\|_{\infty}^{\v}\leq\|\bph\|$, which implies that~\eqref{Theorem:3-1} holds for $k=0$. Assume that the induction hypothesis holds for all $k$ up to some $m$, \textit{i.e.}, $x(k)\leq \gamma^k\|\bph\|\v $ for $k=1,\ldots,m$. Since $f$ and $g$ are homogeneous and order-preserving, it follows that
\begin{align}
\begin{split}
\f\bigl(\x(m)\bigr)&\leq \gamma^m\|\bph\| \f\bigl(\v\bigr),\\
\g\bigl(\x(m-d(m))\bigr)&\leq \gamma^{m-d_{\max}}\|\bph\| \g\bigl(\v\bigr),
\end{split}
\label{Inequality:5.1}
\end{align}
where we used the fact that $\gamma<1$ and $d(m)\leq d_{\max}$ to get the second inequality. Using~\eqref{Theorem:5-0} and~\eqref{Inequality:5.1}, we obtain
\begin{align*}
\frac{1}{v_i}x_i\bigl(m+1\bigr)&= \frac{1}{v_i}\bigl(f_i(\x(m))+g_i(\x(m-d(m)))\bigr)\\
&\leq \gamma^m\|\bph\|\left(\left(\frac{f_i(\v)}{v_i}\right)+\left(\frac{g_i(\v)}{v_i}\right)\gamma^{-d_{\max}}\right)\\
&\leq\gamma^{m+1}\|\bph\|,\quad i=1,\ldots,n.
\end{align*}
It follows from the definition of weighted $l_{\infty}$ norm that $\| \x(m+1)\|_{\infty}^{\v} \leq  \gamma^{m+1} \|\bph\|$. This completes the induction proof.

$(b)\Rightarrow (a):$ Suppose~\eqref{System:3} is globally exponentially stable for all bounded delays. Particularly, let~$d(k)=0$. Then, system $\x\bigl(k+1\bigr)=\f\bigl(\x(k)\bigr)+\g\bigl(\x(k)\bigr)$ is globally asymptotically stable. Since $f+g$ is continuous, order-preserving, and $(\f+\g)(\bnull)=\bnull$ , the conclusion follows from~\cite[Propositions 5.2 and 5.6]{Dashkovski:06}.

\subsection{Proof of Theorem~\ref{Theorem:6}}\label{appendix:4}

We use perfect induction to prove that the desired relation~\eqref{Theorem:6-2} holds. For each $i$, we have
\begin{align*}
\frac{1}{v_i}\bigl|x_i(k+1)\bigr|&=\frac{1}{v_i}\biggl|\sum_{j=1}^n a_{ij} x_j(k)+\sum_{j=1}^n b_{ij} x_j\bigl(k-d(k)\bigr)\biggr|\nonumber\\
&\leq \frac{1}{v_i}\biggl\{\sum_{j=1}^n \bigl|a_{ij}\bigr| \bigl|x_j(k)\bigr|+\sum_{j=1}^n \bigl|b_{ij}\bigr| \bigl|x_j\bigl(k-d(k)\bigr)\bigr|\biggr\}.
\end{align*}
On the other hand, by~\eqref{Theorem:6-0}, Equation~\eqref{Theorem:6-1} always admits a unique solution $\gamma_i\in(0,1)$ for each~$i$. Let $\gamma=\max_{1\leq i\leq n} \gamma_i$. It follows that
\begin{align*}
\biggl(\sum_{j=1}^n \frac{1}{v_i}\bigl|a_{ij}|v_j\biggr)+\biggl(\sum_{j=1}^n \frac{1}{v_i}\bigl|b_{ij}|v_j\biggr)\gamma^{-d_{\max}}\leq \gamma,
\quad i=1,\ldots,n.
\end{align*}
The rest of the proof is similar to the proof of Theorem~\ref{Theorem:4} and is thus omitted.

\bibliographystyle{IEEEtran}
\bibliography{bibliografia}

\end{document}